# Berry Curvature and Topological Hall Effect in Magnetic Nanoparticles

Ahsan Ullah*, Balamurugan Balasubramanian, Bibek Tiwari, Bharat Giri , David J. Sellmyer‡, Ralph Skomski‡, Xiaoshan Xu*

*Department of Physics & Astronomy and Nebraska Center for Materials and Nanoscience, University of Nebraska, Lincoln, NE 68588*

*E-mail: aullah@huskers.unl.edu, xiaoshan.xu@unl.edu

‡ Deceased

***Abstract***: Analytical calculations and micromagnetic simulations are used to determine the Berry curvature and topological Hall effect (THE) due to conduction electrons in small ferromagnetic particles. Our focus is on small particles of nonellipsoidal shapes, where noncoplanar spin structures yield a nonzero topological Hall signal quantified by the skyrmion number $Q$. We consider two mechanisms leading to noncoplanarity in aligned nanoparticles, namely flower-state spin configurations due to stray fields near corners and edges, and curling-type magnetostatic selfinteractions. In very small particles, the reverse magnetic fields enhance $Q$ due to the flower state until the reversal occurs, whereas for particles with a radius greater than coherence radius $R_{\mathrm{coh}}$ the $Q$ jumps to a larger value at the nucleation field representing the transition from the flower state to the curling state. We calculate the Skyrmion density (average Berry curvature) from these spin structures as a function of particle size and applied magnetic field. Our simulation results agree with analytical calculations for both flower state and flux closure states. We showed the presence of Berry curvature in small particles as long as the size of the particle is less than the single domain limit. Using magnetic force microscopy (MFM), we also showed that in a nanodot of Co with a suitable size, a magnetic vortex state with perpendicular (turned-up) magnetization at the core is realized which can be manifested for Berry curvature and emergent magnetic field in confined geometries for single domain state at room temperature.

# I. Introduction

When a quantum system traverses in a continuous parameter space slowly, it follows the environment adiabatically, i.e., remaining in the same eigenstate according to the local value of the parameters. Consequently, the wavefunction of the system accumulates a non-dynamic phase called the Berry phase, which is equal to the path integral of Berry connection in the parameter space, or the areal integral of Berry curvature in case the trajectory is a closed loop [1, 2]. This mechanism applies to the process of itinerant electrons flowing through a spin texture in real space, where the spin of the itinerant electron follows the direction of the local spin [3]. The change of spin direction of the itinerant electrons can be described using an emerging magnetic field $\vec{B}_e$ which is proportional to the Berry curvature $\vec{\Omega}$ as $\vec{B}_e = \frac{\hbar}{e}\vec{\Omega}$, where $\hbar$ and $e$ are reduced Plank constant and electronic charge [4]. The Berry curvature can be found from the spin texture

$$\vec{\Omega} = -\epsilon_{ijk}\frac{1}{2}\boldsymbol{S}\cdot\left(\partial_i\boldsymbol{S}\times\partial_j\boldsymbol{S}\right), \tag{1}$$

where $\boldsymbol{S}(\boldsymbol{r})$ is the unit vector describing the spin direction at position $\boldsymbol{r}$, $\epsilon_{ijk}$ is the antisymmetric tensor. This can be understood using Fig. 1. When $\boldsymbol{S}\cdot\left(\partial_i\boldsymbol{S}\times\partial_j\boldsymbol{S}\right)\neq 0$, the electron trajectory can enclose a non-zero solid angle in the Bloch sphere, which is equal to the Berry phase accumulated.

A famous example that leads to non-zero $\vec{\Omega}$ is the Skyrmion, which is a two-dimensional (2D) spin texture of rotational symmetry [3,4] whose $\boldsymbol{S}$ rotates ±180º from the center to the boundary. In general, the areal integration of $\vec{\Omega}$ of a 2D spin texture of rotational symmetry follows $Q = \frac{1}{2\pi}\int\vec{\Omega}\cdot d\vec{A} = S_z(\infty) - S_z(0)$, where $S_z(0)$ and $S_z(\infty)$ are the z component of $\boldsymbol{S}$ at the center and far away. $Q$ is also called Skyrmion number because the value is ±1 for Skyrmion [4]. Correspondingly, the areal integral of the emerging field of the Skyrmion is $\int\vec{B}_e\cdot d\vec{A} = \pm\frac{h}{e}$ or the magnetic flux quantum. Some other examples of spin texture with finite Skyrmion number are chiral spin texture, magnetic vortex, and meron [5,6].

Given that the areal integral of $\vec{\Omega}$ and $\vec{B}_e$ are constant for a Skyrmion regardless of its area, the average Berry curvature $\vec{\Omega}$ and the emerging field $\vec{B}_e$ are then inversely proportional to the Skyrmion area A, or proportional to the areal density of the Skyrmion Ψ as $<|\vec{\Omega}|> = \frac{2\pi Q}{A} = 2\pi\Psi$. For an area A with multiple Skyrmions, the average $\vec{B}_e$ and $\vec{\Omega}$ only depends on the topology of the spin texture, i.e., the number of Skyrmions $N$, or the number of "holes" in the spin texture as $<|\vec{\Omega}|> = 2\pi\Psi = \frac{2\pi NQ}{A}$, regardless of the shape of the "holes". This is the origin of the name topological Hall effect (THE) generated by $\vec{B}_e$ [4,3,7,8]. The Berry-curvature contributions to the Hall effects, i.e., THE, represent a research topic in its own right but are also technologically interesting.

In this work, we investigate the spin structure in ferromagnetic nanoparticles and the corresponding Berry curvature. Previous work shows that thin films exhibit magnetic skyrmions and other types of spin textures, which leads to an emergent magnetic field and topological Hall effect [4,3,7,8]. Traditionally, such spin textures are not expected in small isolated nanoparticles because the exchange interaction ensures parallel spin alignment on small-length scales [9,10,11]. However, nanogranular materials exhibit features similar to that of thin films with skyrmions-like spin textures [12,13,14,15]. These examples are related to interacting magnetic nanoparticles and grains. Recently, it has been reported that isolated nanoparticles of noncentrosymmetric B-20 structures can exhibit geometrically stabilized skyrmionic spin textures [16]. However, Berry curvature in non-interacting centrosymmetric nanoparticles with minimum magnetic energy is yet to be systematically modeled.

In this work, using a micromagnetic approach the Berry curvature for nanoparticles with nonellipsoidal shapes, which tend to exhibit flower-state spins structures [10, 17] that compete against other spin states, such as vortex states due to magnetostatic flux closure [9,10,11,17,18,19]. Our focus is on the transition between flower and curling states as the size of nanoparticle changes across the coherence radius $R_{coh}$ [9,10,11,20] and the corresponding Berry curvature. $R_{coh}$ reflects the competition between magnetostatic and exchange energies. It is unrelated to the critical single-domain size $R_{SD}$, which can be much larger than $R_{coh}$ [21, 26]. In general flux closure is favorable but competes against the exchange interaction. The latter scales as $A_e/R^2$ and is therefore less important in big particles. In small particles, the $A_e/R^2$ term completely kills the curling/vortex contribution and largely (but not completely) the flower-state contribution and the Berry curvature. We used analytical and computational modeling to study these effects on Berry curvature in nanoparticles.

# II. Methods

Throughout the paper, we express the local magnetization $\boldsymbol{M}(\boldsymbol{r})$ as $\boldsymbol{M}(\boldsymbol{r}) = M_s\, \boldsymbol{S}(\boldsymbol{r})$, where $\boldsymbol{S}(\boldsymbol{r})$ is a unit vector representing the normalized quasi-classical spin. The spin of the charge carriers is assumed to follow the local magnetization adiabatically so that $\boldsymbol{S}(\boldsymbol{r})$ also characterizes the conduction electrons. These spin structures are investigated analytically, by considering an approximate Hamiltonian, and by micromagnetic simulation using *Ubermag* [22]. The determination of local configuration $\boldsymbol{M}(\mathbf{r}) = M_s\, \boldsymbol{S}(\boldsymbol{r})$, done by energy function is given by [11, 23]:

$$\mathscr{E} = \int \left\{ A_e \left[\nabla\left(\frac{\boldsymbol{M}}{M_s}\right)\right]^2 - K_1 \frac{(\boldsymbol{n}.\boldsymbol{M})^2}{M_s^2} - \mu_o\, \boldsymbol{M}.\boldsymbol{H} - \frac{\mu_o}{2} \boldsymbol{M}.\boldsymbol{H}_d(\boldsymbol{M}) \right\} dV \qquad (2)$$

Where $A_e$ is the exchange stiffness, $K_1(\boldsymbol{r})$ denotes the first uniaxial anisotropy constant, $\boldsymbol{n}(\mathbf{r})$ is the unit vector of the local anisotropy direction, $\boldsymbol{H}$ is the applied field and $\boldsymbol{H}_d(\boldsymbol{M})$ is the magnetostatic self-interaction. Physically $\boldsymbol{M}(\mathbf{r})$ corresponds to local and global minima. In this research article, we provide an analytical scaling analysis to gauge the broad range of effects important in the present context.

To model the spin textures in nanoparticles and study Berry-phase numerically, we have performed micromagnetic simulations using *ubermag* supported by OOMMF [22]. We have numerically extracted the skyrmion number $Q$ from the spin structure. For simulations different sizes of nanoparticles were considered for the study of flower, curling, and vortex state. We have used a computational cell size of 1.9 nm, which is well below the exchange length $l_{ex}$ [11].

Samples of Co ferromagnetic dots were made by using electron-beam lithography and evaporation in an ultrahigh vacuum using an electron-beam gun. The circular nanodot patterns were defined on thermally oxidized Si substrates with positive resists. The bilayer positive resists PMMA950/MMA EL6 was exposed to an electron beam and the liftoff method was used to create the circular pattern. Ordered Co circular arrays were fabricated, in a trilayered structure of $Ti/Co/SiO_2$ were fabricated. The circular layered structure was grown by e-beam evaporation in a UHV system. The base pressure was in the range of $1\times10^{-8}$ torr. The evaporation pressure is less than $5\times10^{-7}$ torr. The thickness is in the range of ~20 nm for Ti and ~40 nm for Co, and ~20 nm $SiO_2$ layer to prevent oxidation. The thickness was monitored during growth by a quartz balance for each layer. By a lift-off process, the resist is removed and dots with designed sizes remain on top of the Si surface. A Bruker Dimension Icon® Atomic Force Microscope was used to map the topography and magnetic images at room temperature. During the measurement, magnetic force microscopy was performed in constant height mode (single pass).

# III. Calculations and Results

Equation (1) means that the Berry curvature and Skyrmion number are unique functions of the spin structure $\boldsymbol{S}(\boldsymbol{r})$, which is determined by the magnetic interactions and sample geometry. $\boldsymbol{S}(\boldsymbol{r})$ describes the orientation of magnetization at position $\boldsymbol{r}$, which can be written as [20]:

$$\boldsymbol{S}(\boldsymbol{r}) = \sin\Theta(\boldsymbol{r})\cos\Phi(\boldsymbol{r})\,\boldsymbol{e}_x + \sin\Theta(\boldsymbol{r})\cos\Phi(\boldsymbol{r})\,\boldsymbol{e}_y + \cos\Theta(\boldsymbol{r})\,\boldsymbol{e}_z, \quad (3)$$

Where $\boldsymbol{e}_x$, $\boldsymbol{e}_y$, and $\boldsymbol{e}_z$ are unit vectors along the x, y, and z directions respectively, $\Theta(\boldsymbol{r})$ and $\Phi(\boldsymbol{r})$ are the polar and azimuthal angles. Both the flower state and the curling state are small deviations from the homogeneous magnetization along the easy axis direction (z-axis). The deviation, or the x-y component of $\boldsymbol{S}$

$$\boldsymbol{S}_{xy}(\boldsymbol{r}) = \sin\Theta(\boldsymbol{r})\cos\Phi(\boldsymbol{r})\,\boldsymbol{e}_x + \sin\Theta(\boldsymbol{r})\cos\Phi(\boldsymbol{r})\,\boldsymbol{e}_y \quad (4)$$

is along the radial direction and the azimuthal direction for the flower and the curling state (see Fig. 2) respectively. Using cylindrical coordinates, i.e., $\boldsymbol{r} = (\rho, \varphi, z)$, one can describe the flower state and the curling state using $\Phi(\boldsymbol{r}) = \varphi$ and $\Phi(\boldsymbol{r}) = \varphi \pm \pi/2$ respectively.

Figure 2 shows schematic spin structures encountered in nanoparticles of various cross-sections, which we considered in this paper to study the Emergent magnetic field in these small particles. All structures in Fig. 2 exhibit axial symmetry, that is, the magnetic anisotropy is of the aligned *c*-axis type, and $C_3$, $C_4$, or $C_\infty$ rotations reproduce the original spin structure. Beyond Fig. 1(b), one needs to consider the case of the coherent state, where $\boldsymbol{S}(\boldsymbol{r})$ is almost constant in the particle and forms an angle with the symmetry axis of the particle i.e. flower state.

## A. Flower State

In particles smaller than about 10 nm, the magnetization $\boldsymbol{M}(\boldsymbol{r})$ is almost uniform, as in Fig. 2(b-c). The reason is the interatomic exchange described by the energy density $A_e\,(\nabla\boldsymbol{S})^2$, where $A_e \sim 10^{-11}$ pJ/m is the exchange stiffness. The gradient term suppresses the magnetization inhomogeneities and scales as $1/R^2$, where $R$ is the particle radius. Particles having radii of several 10 nm tend to exhibit nonuniform spin configurations, such as the side view in Fig. 1(b) and the top views shown in Figs. 2(e-f). The non-uniform state is called the flower state in which the spins $\boldsymbol{S}(\boldsymbol{r})$ close to the edge, forms an angle $\Theta(\boldsymbol{r})$ with the symmetry axis of the particle. The flower state is limited to non-spherical shapes, ideally cubes. In very small nanoparticles, the exchange energy $\sim A_e/R^2$ dominates and $\Theta(\boldsymbol{r})$ approaches zero as shown in Fig. 2(b, c). When the size increases the 'flower opens', i.e., the magnetization on these edges rotates away from the parallel orientation [17, 26]. The flower-state spin structure is well established and used as a standard problem in micromagnetism [18, 17, 26, 28]. In the flower state, spins tilt away from the z-axis with 'radial' symmetry. The tilt angle increases with the distance from the center, as illustrated using the spin at the particle edges or corners in Fig. 2(d-f). The angle $\Theta$ not only depends on the dimensions of the nanoparticles but also on the external magnetic field. We used micromagnetics to find $\Theta(\boldsymbol{r})$ as a function of particle size and external magnetic field. The spin structure $\boldsymbol{S}(\boldsymbol{r})$ is determined by minimizing the micromagnetic (free) energy:

$$\mathscr{E} = \int \eta \, \mathrm{d}V \quad (5)$$

Here the energy density $\eta$ contains exchange, anisotropy, Zeeman, and magnetostatic-selfinteraction for flower-state contribution as described in Eq. (2). Here we show analytical calculations based on a set of simplifications to demonstrate the physical picture in a semi-quantitative way. We perform an approximate volume-averaging first so that $\int \eta \, dV = \langle\eta\rangle V$. and then minimize $\langle\eta\rangle$ with respect to Θ, which is the polar angle of the magnetization at the particle edges or corners (blue arrows in Fig. 2); sinΘ is the length of the blue arrows in Fig. 2(d). An exact determination of the average $\langle\eta\rangle$ is highly nontrivial because it requires the knowledge of $\boldsymbol{S}(\boldsymbol{r})$. However, $\boldsymbol{S}(\boldsymbol{r})$ is subject to some constraints (normalization and symmetry) and is approximately known for several cases. To determine the parameters in Eq. (3), we use the approximation of small values Θ. With the assumption of small Θ and find [see supplementary material eq. S11-13]:

$$\langle\eta\rangle = \left(\frac{A_e}{R^2} + \left(K + \frac{\mu_o}{2} M_s H\right)\right) (\sin^2\Theta) - \mu_o M_s H_F \sin\Theta. \tag{6}$$

As usual, the exchange stiffness $A_e$ parameterizes the interatomic exchange, $\eta_A = A_e(\nabla \boldsymbol{S})^2$, whereas $K$ is the uniaxial anisotropy of the particles, which are assumed to be $c$-axis aligned in the z-direction. This $K$ includes both magnetocrystalline ($K_1$) and shape-anisotropy contributions. There are three magnetostatic terms in Eq. (6), namely the Zeeman interactions with the external magnetic field $H$, the magnetostatic selfinteraction energy described by the demagnetizing factor $D$ which gives the flux closure, and the flower-state energy correction ($H_F$) due to the nonuniform magnetization inside the particles. The spin texture due to flux closure, i.e., the curling state is absent in small particles ($R < R_{coh}$). Overall, the flower state reduces the selfinteraction energy compared with the homogeneous state, which is accounted for using an energy correction $\mu_o M_s H_F \sin\Theta$, where $H_F$ is assumed a positive constant. The field $H_F$ parameterizes the interaction of the nanoparticle spins with the in-plane component of the demagnetizing field inside the nanoparticles. This field is zero for homogeneously magnetized ellipsoids but nonzero for magnetized particles of arbitrary shape (see Fig. 2) where it gives rise to flower-state spin structures such as those in Figs. 2(d-f). The parameter $H_F$ depends on the shape of the particles, especially on the cross-section, but is generally comparable to but somewhat smaller than the saturation magnetization $M_s$. To find the stable state, we minimize $\langle\eta\rangle$ using Eq. (6) with respect to Θ. Equation (6) is quadratic in sinΘ, and therefore easily minimized. Explicitly, one has:

$$\sin\Theta = \frac{\mu_o M_s H_F}{\frac{A_e}{R^2} + 2K + \mu_o M_s H}. \tag{7}$$

In Eq. (7), $H_F$ is positive, meaning that the magnetization has a component pointing away from the symmetry axis. Essentially, a larger Θ increases the exchange energy and anisotropy energy but decreases the Zeeman energy and the selfinteraction energy. It is also clear that when external field $H$ is large enough, sinΘ diverges, corresponding to the magnetization reversal.

The Berry curvature can be calculated accordingly using Eq. (1) using $\boldsymbol{S}(\boldsymbol{r})$. The integration of Berry curvature over the entire magnetic particles gives the Skyrmion number. It can be shown that the Skyrmion number for a cylindrical particle is:

$$Q = \frac{\cos\Theta_c - \cos\Theta_\infty}{2} \tag{8}$$

where $\Theta_c$ and $\Theta_\infty$ are the polar angles of $\boldsymbol{S}$ at the center and far away. In the flower state discussed here, $\Theta_c = 0$ and $\Theta_\infty = \Theta$. One can calculate the skyrmion number from Θ in Eq. (7), which clearly shows that $\Theta \to 0$ when $R \to 0$. Eq. (7) also shows that $\Theta$ saturates when $R \to \infty$. Microscopically, since the exchange term ($A_e$) is proportional to $\sin^2\Theta$, it tends to suppress the magnetization gradients (spin tilt). This suppression is most effective in small particles since the exchange energy is proportional to $\frac{A_e}{R^2}$.

We plotted the analytical calculations for skyrmion number as a function of particle size at zero magnetic fields and as a function of magnetic field for a particle of size 10 nm. Figure 3(a) shows the change in skyrmion number as a function of particle size. The striking feature in Fig. 3(a) is that the skyrmion number is almost zero for very small particles, rapidly increases at a certain size, and slowly converges to a limiting curve representing the size where the flower state does not exist and curling due to self-interaction appears. While the corresponding transition size depends on the micromagnetic parameters, the overall trend can be qualitatively understood using Eq. (7). Figure 3(b) shows the skyrmion number as a function of the magnetic field. The high-field $Q$ is very small but nonzero, except for aligned ellipsoids where $Q = 0$ due to the absence of edges or corners. In principle, the high field state is still a flower state like those in Fig. 2(d-f), but $\Theta$ is so small that they are close to uniform magnetization shown in Fig. 2(b-c). In Fig 3(b) the skyrmion number achieves the maximum value just before the magnetization reversal representing the maximum opening of the flower under the reverse field. Positive and negative magnetic fields align and misalign the magnetization, respectively, and the latter, which amounts to increasing lengths of the blue arrows in (d-f), is plotted in Fig. 3(b).

Fig. 4 shows the vector 3d plot for the flower state using Eq. 3 with $\Phi(\boldsymbol{r}) = \varphi$ [Fig. 4(a)] and emergent magnetic field [Fig. 4(b)] due to the flower state in cylindrical nanoparticles. The emergent magnetic field which is proportional to the Berry curvature gives rise to THE. Figures 3 and 4 have important implications for the experimental and numerical investigation of THE in nanoparticles with dimensions lower than coherent sizes.

## B. Magnetization Curling

In reality, interatomic exchange ($A_e$) favors magnetization uniformity but competes against other energy contributions, such as magnetostatic energies. Magnetostatic interactions favor flux closure ($\nabla\cdot\boldsymbol{M} = 0$) over magnetic poles ($\nabla\cdot\boldsymbol{M} \neq 0$) [24, 25], and this principle manifests in the existence of a curling mode during nucleation and with the increase of the size of nanoparticles [11, 24, 26]. This flux closure reduces the magnetostatic energy and therefore yields a nonzero azimuthal component to magnetization to the perpendicular component [24]. Therefore, magnetization reversal in ellipsoids and cylinders having radii larger than coherent radius $R_{coh}$ leads to magnetization curling and then vortex as long as the size in the single domain limit ($R_{coh} < R < R_{SD}$) [26, 27] and the energy correction ($H_F$) giving flower state is absent. Curling is a special vortex state, occurring in the intermediate vicinity of the nucleation field during the magnetization reversal and $R_{coh}$. It is symmetric around the z-axis. The curling mode, which has the symmetry of Fig. 2(g-h), is one of the few exact solutions in many-body physics [10, 23]. The question, therefore, is how the curling mode yields a topological Hall effect and how this effect depends on the particle. Our interest is to study the skyrmion density at the nucleation field and particles with a radius slightly greater than $R_{coh}$ and much lower than $R_{SD}$.

Again, the curling mode is defined by small perpendicular magnetization components $|\boldsymbol{S}_{xy}(\boldsymbol{r})| \ll 1$. As the size of nanoparticles or the reverse magnetic field increase, $|\boldsymbol{S}_{xy}(\boldsymbol{r})|$ increases, and the magnetic state eventually becomes the curling state. Due to the flux closure, the curling state does not require the correction to the selfinteraction energy as introduced in the flower state,

i.e., $H_F$=0 and the demagnetization field $\mathbf{H}_d(\mathbf{M}(r))$ provides the magnetostatic selfinteraction for flux closure. The equation of state can be obtained by minimizing the total energy eq. (2) with respect to $S_{xy}(\boldsymbol{r})$ [10] (see supplement eq. S(17-21)):

$$(2A_e\nabla^2 - 2K_1 - \mu_0 M_s H + \mu_0 D M_s^2)\boldsymbol{S_{xy}}(\boldsymbol{r}) = 0 \quad (9)$$

In Eq. (9), $\boldsymbol{H}_d = -D\mathbf{M}$ where $D$ is the demagnetization factor addition giving flux closure. The curling mode considered by Frei [20] can be written by using a small-angle approximation. Brown considered the exact solution of a specific micromagnetic problem under the cylindrical symmetry, i.e., $S_{xy}(\boldsymbol{r}) = S_{xy}(\rho)$ [9, 10]. Recall in the curling state, $\Phi(\boldsymbol{r}) = \varphi \pm \pi/2$, Eq. (3) can be rewritten as:

$$M(r) = M_s\left[-S_{xy}(\rho)\sin(\phi)\,\boldsymbol{e_x} + S_{xy}(\rho)\cos(\phi)\,\boldsymbol{e_y} + S_z\boldsymbol{e_z}\right] \quad (10)$$

where $S_{xy} = \sin\Theta$ and $S_z = \cos\Theta$. Because it is Eigen function of differential eq. (9), substituting the curling mode considered in Eq. (10) in micromagnetic Eq. (9) leads to:

$$\left(\rho^2\frac{\partial^2}{\partial\rho^2} + \rho\frac{\partial}{\partial\rho} + ((\mathrm{k}\rho)^2 - 1)\right)S_{xy}(\rho) = 0 \quad (11)$$

where $k^2 = -\left(\frac{2K_1 + \mu_0 M_s H - \mu_0 M_s^2 D}{2A_e}\right)$. Eq. (10) is the Bessel equation and therefore, we can write the curling mode in the cylinder $S_{xy}(\rho) = J_1(k\rho)$ or $\sin\Theta \approx \Theta(\rho) = J_1(k\rho)$. The boundary condition is modified as $\frac{\partial J_1(k\rho)}{\partial\rho}|_{\rho=R} = 0$. This boundary condition is possible only when $kR = q_1 = 1.841$. In addition, in spherical particles with $D = 1/3$, the curling mode can be described by a spherical Bessel function, $S_{xy}(\rho) = j_1(k\rho)$ with the smallest root $kR = q_2 = 2.0816$. The skyrmion density due to magnetization curling in cylinder and sphere calculated by using Eq. 1 as:

$$\Psi_{cylinder} = \frac{J_1(k\rho)}{\rho}\frac{\partial J_1(k\rho)}{\partial\rho} \quad (13)$$

$$\Psi_{sphere} = \frac{j_1(k'\mathrm{r})}{\mathrm{r}}\frac{\partial j_1(k'\mathrm{r})}{\partial\mathrm{r}}. \quad (14)$$

Figures 5(a) and 6(a) show the curling mode vector 3d plot in a cylinder using $S_{xy}(\rho) = J_1(k\rho)$ and $\Phi(\boldsymbol{r}) = \varphi \pm \pi/2$ in Eq. (3) . Figure 6(b) shows the emergent magnetic field within the cylinder due to curling mode. The skyrmion number is obtained by integrating the Berry curvature in limits of 0 to $R$, where $R$ is the radius of the cylinder or sphere provided that $R > R_{coh}$:

$$Q_{\mathrm{cylinder}} = \frac{1}{2}J_1(kR)^2 \quad (15)$$

$$Q_{\mathrm{sphere}} = \frac{1}{2}j_1(kR)^2. \quad (16)$$

In a high magnetic field, the particle will show a flower state with a small $\Theta$, which corresponds to a small skyrmion number or emergent magnetic field. At the nucleation field, $H$ =

$H_n$, the flower state turns into a curling state. The skyrmion number and hence the emergent field and THE will jump to a large value. For particle $R < R_{coh}$ with only flower state, we do not have any nucleation field so in that case we do not have any jump change in the skyrmion number. These observations are also shown in micromagnetic simulations in the next section. The nucleation field at which the curling appears is given by [23]:

$$H_n = \frac{2K_1}{\mu_0 M_s} - DM_s + \frac{2A_e q_i^2}{\mu_0 M_s R^2} \tag{17}$$

where $q_1 = 1.841$, $D = 0$ for the cylinder and $q_2 = 2.0816$ , $D = 1/3$ for the sphere. In a cylinder of radius "$R$" the boundary condition gives $kR = q_1 = 1.841$ and $Q_{\text{cylinder}} = 0.17$; for a sphere $k'R = q_2 = 2.0816$ and $Q_{\text{sphere}} = 0.11$. This skyrmion number will increase as a function of the field after the nucleation field until there is a magnetization reversal in the nanoparticle. For the fixed value of $R > R_{coh}$ where curling appears $Q$ has a constant value. It's because the curling mode is subject to the eigenvalue condition $kR = 1.84$ so $Q = 0.5J_1(1.84)^2 = 0.17$. This also applies to the following two considerations. The $J_1(kr)$ oscillations describe radial spin waves. The curling mode is a 1s state in the analogy of an electron in a cylinder, where the lowest-lying excited radial spin-wave mode is a 2s state and has $kR = 5.33$ and Q = 0.06 showing the Berry curvature of electron scattering from excited states is tricky from the viewpoint of dynamics [23].

For larger values of radii $R$, $S_{xy}$ becomes larger and eventually, the curling mode turns into a vortex mode. As the size of the nanoparticle increases the long-range magnetostatic interaction between $\boldsymbol{M}(\mathbf{r})$ and $\boldsymbol{M}(\mathbf{r}')$ gives rise to in-plane spin configurations. Due to the increase in size, the magnetostatic self-interaction has a higher magnitude compared with the short-range exchange interactions.

## C. Micromagnetic Simulations

The micromagnetic simulations are performed near coherent radius $R_{coh}$ using OOMMF-based software *ubermag* [22]. The parameters which we used for simulations are $M_s=1.2\times10^6$A/m, $A_e=1\times10^{-11}$J/m, $K_1=0.2$MJ/m$^3$, $l_o = (A_e/\mu_o M_s^2)^{1/2}$. In all cases, cell size is considered smaller than the exchange length. For simulation, we considered ferromagnetic cubes, cylinders, and spheres with uniaxial anisotropy. We changed the radii of the cylinder and sphere and simulated the skyrmion number for different radii. Our results show that as the size increased the skyrmion number $Q$ for the flower state also increases representing an increase in the angle $\Theta$. We calculated the skyrmion number for a size lower than critical single-domain size i.e., $R < R_{coh}$. The simulation results show that the flower state only exists in either cylinder or a ferromagnetic cube. In the sphere, the flower state does not open due to a small edge effect i.e., $H_F = 0$.

Figure 7 shows the Skyrmion number hysteresis and normalized magnetization as a function of the decreasing magnetic field $B(T)$ in a cube of $L < L_{coh}$. This indicates the presence of a flower state, which appears during magnetization reversal. At a higher field, the spin aligns in the direction of the magnetic field showing a lower value of the Skyrmion number. At the maximum value of the skyrmion number, the magnetization is some fraction of $\pm M_s$ (which is the minimum value of magnetization just before reversal). The skyrmion number $Q$ changes the sign at the reversal indicating the change of magnetization $+M_z$ to $-M_z$ as the applied field changes from $B_{\max}$ to -$B_{\max}$. $Q$ achieves maximum value just before the reversal (the point where we have a minimum value of magnetization and maximum opening of flower state) close to the coercivity field.

As the nanoparticles size increases, i.e., $R > R_{coh}$, the magnetization reversal involves the presence of curling mode. Micromagnetic simulation for cylinder (Fig 8(a)) and sphere (Fig. 8(b)) for radius $R > R_{coh}$ shows the presence of curling which appears during magnetization reversal. For cylinders and cubes, the spin texture first involves the presence of a flower state as the spin aligns in the direction of the strong external field except for the tilted spins at the edges. At the nucleation field the spin texture change to a curling state with core polarity in p = +1. At the magnetization reversal, the polarity changes to p = -1 with curling mode. At the high field, the curling mode disappears and again gives rise to a flower state with polarity pointing in the direction of magnetization. In other words, we can say that in nanoparticles with non-coplanar and noncollinear spin texture the core polarity is directed in the direction of magnetization. The absence of THE contribution in the sphere at a high magnetic field indicates the absence of a flower state in the sphere as compared with the cylinder which has a finite skyrmion number at high fields indicating the effect of edges i.e. $H_F \neq 0$.

## D. Experimental Studies of Magnetic Vortex

Based on our calculation, it is evident that the single domain state gives rise to the finite emergent magnetic field and THE. Additionally, our micromagnetic experiment Fig. 8(a) and Fig. S1 shows that the flower and curling state appear at higher and lower fields respectively as long as the radius of the particle satisfies $R_{\rm coh} < R < R_{\rm SD}$. To study the single domain states we carried out magnetic force microscopy (MFM) measurements on circular dots of Co that give evidence for the presence of a vortex state with a perpendicular magnetization core.

In MFM, the instrument was operated in ac mode to detect the magnetic force acting between the cantilever tip and the surface of the circular dots in ambient conditions. To minimize the stray-field effect, the low-moment CoCr tip was used. The distance between the surface and the tip was between 30 nm. The MFM and AFM images are shown in Figs. 9&10. The MFM images for most of the circular nanodots show a clear contrast between the center and the surrounding. The spins in the dots align parallel to the plane and at the dark spot, the spin aligns perpendicular to the plane as shown in Fig. 9(b). The area of the dark region is very small for vortex in permalloy [19, 29] but the anisotropy of Co is large making it a semi-hard magnet. This gives rise to a significantly large area for the dark spot in the middle of the single-domain magnetic state. These spin configuration in nanodots appears when the dot thickness becomes much smaller than the dot diameter and all spins tend to align in-plane giving a vortex. Our micromagnetic simulation Fig. 9(c) for the Co nanodisk also shows that the Co disk of up to 500 nm diameter exhibits a magnetic vortex state.

Fig. 8(a) shows that particles with radius $R_{\rm coh} < R < R_{\rm SD}$, where $R_{\rm SD}$ is the radius of a single domain, constitute two states at different magnetic fields. At a very high field the presence of finite skyrmion number is due to flower states, while at a low field, it represents the curling/vortex-like spin textures. So, we also applied an external magnetic field of 1-tesla perpendicular to the plane using permanent magnets. The image in Fig. S2 shows that the vortex-like state disappears and most of the spin aligns in the direction of the field except at the edges where the spins are slightly tilted showing the edge effects of the flower state for the spins.

## IV. Conclusions

Berry curvature in ferromagnetic nanoparticles is investigated by analytical calculations and micromagnetic simulations by considering the competition between exchange interaction, anisotropy energy, and magnetostatic energy. Our calculations reveal a complex Berry curvature and topological Hall-effect scenario in the nanoparticles. The distinction is between ellipsoidal (spherical) and nonellipsoidal (cubic and cylindrical) nanoparticles, the latter exhibiting flower-state contributions to the Hall effect. We studied finite Skyrmion density due to flower states and curling states both analytically and using micromagnetic simulations. Very small grains have an approximately uniform magnetization, whereas in somewhat bigger grains flower-state, curling, and vortex state appears which gives rise to finite Berry curvature. These results can be very useful for the study of the Hall effect due to the emergent magnetic field. These contributions can be potentially realized experimentally, for example by embedding individual magnetic particles in a nonmagnetic metallic matrix.

**ACKNOWLEDGMENTS:** This research is primarily supported by National Science Foundation under Grant No. OIA-2044049 (NSF-EQUATE). This work was performed in part in the Nebraska Nanoscale Facility and Nebraska Center for Materials and Nanoscience, which are supported by the National Science Foundation under Award No. ECCS: 2025298. This work also used the Holland Computing Center of the University of Nebraska for performing micromagnetic simulations.

## V. References

[1] M. V. Berry, Proc. R. Soc. Lond. A **392**, 45-57 (1984).

[2] D. Xiao, M.-Ch. Chang, and Q. Niu, "Berry phase effects on electronic properties", Rev. Mod. Phys. **82**, 1959 (2010).

[3] S. Seki and M. Mochizuki, Skyrmions in Magnetic Materials, Springer International, Cham (2016).

[4] N. Nagaosa and Y. Tokura, *Nature Nanotech*. **8**, 899 (2013).

[5] Y. Taguchi, Y. Oohara, H. Yoshizawa, N. Nagaosa, and Y. Tokura, Science **291**, 2573 (2001).

[6] J. Bouaziz, H. Ishida, S. Lounis, and S. Blügel, Phys. Rev. Lett. **126**, 147203 (2021).

[7] N. Kanazawa, M. Kubota, A. Tsukazaki,4 Y. Kozuka, K. S. Takahashi, M. Kawasaki, M. Ichikawa,1 F. Kagawa, and Y. Tokura, Phys. Rev. B **91**, 041122(R) (2015).

[8] A. S. Ahmed, J. Rowland, B. D. Esser, S. R. Dunsiger, D. W. McComb, M. Randeria, and R. K. Kawakami, Phys. Rev. Mater. **2**, 041401(R) (2018).

[9] William Fuller Brown, (Jr.) New York, London J. Wiley, (1963).

[10] A. Aharoni, *Introduction to the Theory of Ferromagnetism*, University Press, Oxford 1996.

[11] R. Skomski, Journal of Physics: Condensed Matter, 15, R841-R896 (2003).

[12] W. Zhang, B. Balasubramanian, A. Ullah, R. Pahari, X. Li, L. Yue, Shah R. Valloppilly, A. Sokolov, R. Skomski, and David J. Sellmyer, Appl. Phys. Lett. **115**, 172404 (2019).

[13] B. Balasubramanian, P. Manchanda, R. Pahari, Z. Chen, W. Zhang, S. R. Valloppilly, X. Li, A. Sarella, L. Yue, A. Ullah, P. Dev, D. A. Muller, R. Skomski, G. C. Hadjipanayis, and D. J. Sellmyer, Phys. Rev. Lett. 124, 057201 (2020).

[14] A. Ullah, X. Li, Y. Jin, R. Pahari, L. Yue, X. Xu, B. Balasubramanian, David J. Sellmyer, and R. Skomski, Phys. Rev. B **106**, 134430 (2022).

[15] R. Pahari, B. Balasubramanian, A. Ullah, P. Manchanda, H. Komuro, R. Streubel, C. Klewe, S. R. Valloppilly, P. Shafer, P. Dev, R. Skomski, and D. J. Sellmyer Phys. Rev. Materials **5**, 124418 (2021).

[16] K. Niitsu, Y. Liu, A. C. Booth, X. Yu, N. Mathur, M. J. Stolt, D. Shindo, S. Jin, J. Zang, N. Nagaosa and Y. Tokura, Nature Materials **21**, 305–310 (2022).

[17] W. Rave, K. Fabian, A. Hubert JMMM **190** 332-348 (1998).

[18] M. E. Schabes and H.N. Bertram J. Appl. Phys. **64**, 1347 (1988).

[19] T. Shinjo, T. Okuno, R. Hassdorf, K. Shigeto, and T. Ono, Magnetic vortex core observation in circular dots of permalloy, Science **289**, 930 (2000).

[20] E. H. Frei, S. Shtrikman, and D. Treves Phys. Rev. **106**, 446 (1957).

[21] R. Skomski, H.-P. Oepen, and J. Kirschner, "Micromagnetics of ultrathin films with perpendicular magnetic anisotropy", Phys. Rev. B **58**, 3223-3227 (1998).

[22] M. Beg, M. Lang, and H. Fangohr, IEEE Trans. Magn. 58, 7300205 (2022).

[23] R. Skomski, J. P. Liu and D. J. Sellmyer, Phys. Rev. B **60,** 7359 (1999).

[24] H. Kronmüller and M. Fähnle, Micromagnetism and the Microstructure of Ferromagnetic Solids (University Press, Cambridge 2003).

[25] R. Skomski, Simple Models of Magnetism (University Press, Oxford, 2008).

[26] E. Pinilla-Cienfuegos, S. Mañas-Valero, A. Forment-Aliaga, and E. Coronado, ACS Nano **10**, 1764 (2016).

[27] Sara A. Majetich, Tianlong Wen, and O. Thompson Mefford MRS Bulletin volume **38**, 899–903 (2013).

[28] C. Gatel, F. J. Bonilla, A. Meffre, E. Snoeck, B. Warot-Fonrose, B. Chaudret, L.-M. Lacroix, and T. Blon, Nano Lett. **15**, 6952 (2015).

[29] X. H. Zhang, T. R. Gao, L. Fang, S. Fackler, J. A. Borchers, B. J. Kirby, B. B. Maranville, S. E. Lofland, A. T. N'Diaye, E. Arenholz, A. Ullah, J. Cui, R. Skomski, I. Takeuchi J. Magn. Magn. Mater. 560, 169627 (2022).

[30] R. Skomski, B. Balasubramanian, A. Ullah, C. Binek, and D. J. Sellmyer, AIP Adv. **12**, 035341 (2022).

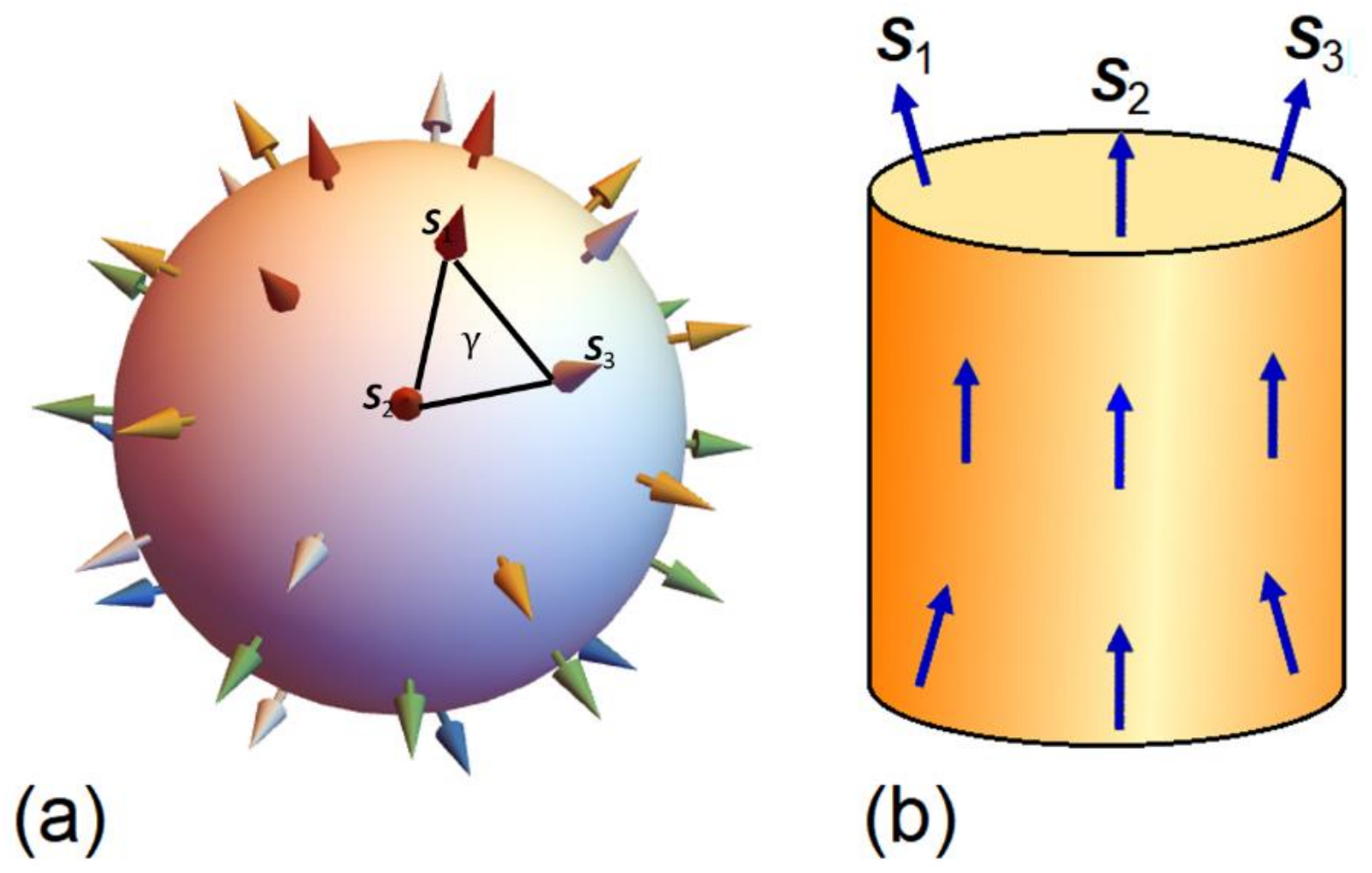

**Fig. 1.** Berry curvature: (a) rotating spin on a Bloch sphere and (b) noncoplanar spins in a nanostructure. When the three vectors are symmetrically arranged and form an angle Θ with the symmetry axis, gives rise to finite spin chirality $\chi_s = \boldsymbol{S}_1 \cdot (\boldsymbol{S}_2 \times \boldsymbol{S}_3)$ [5].

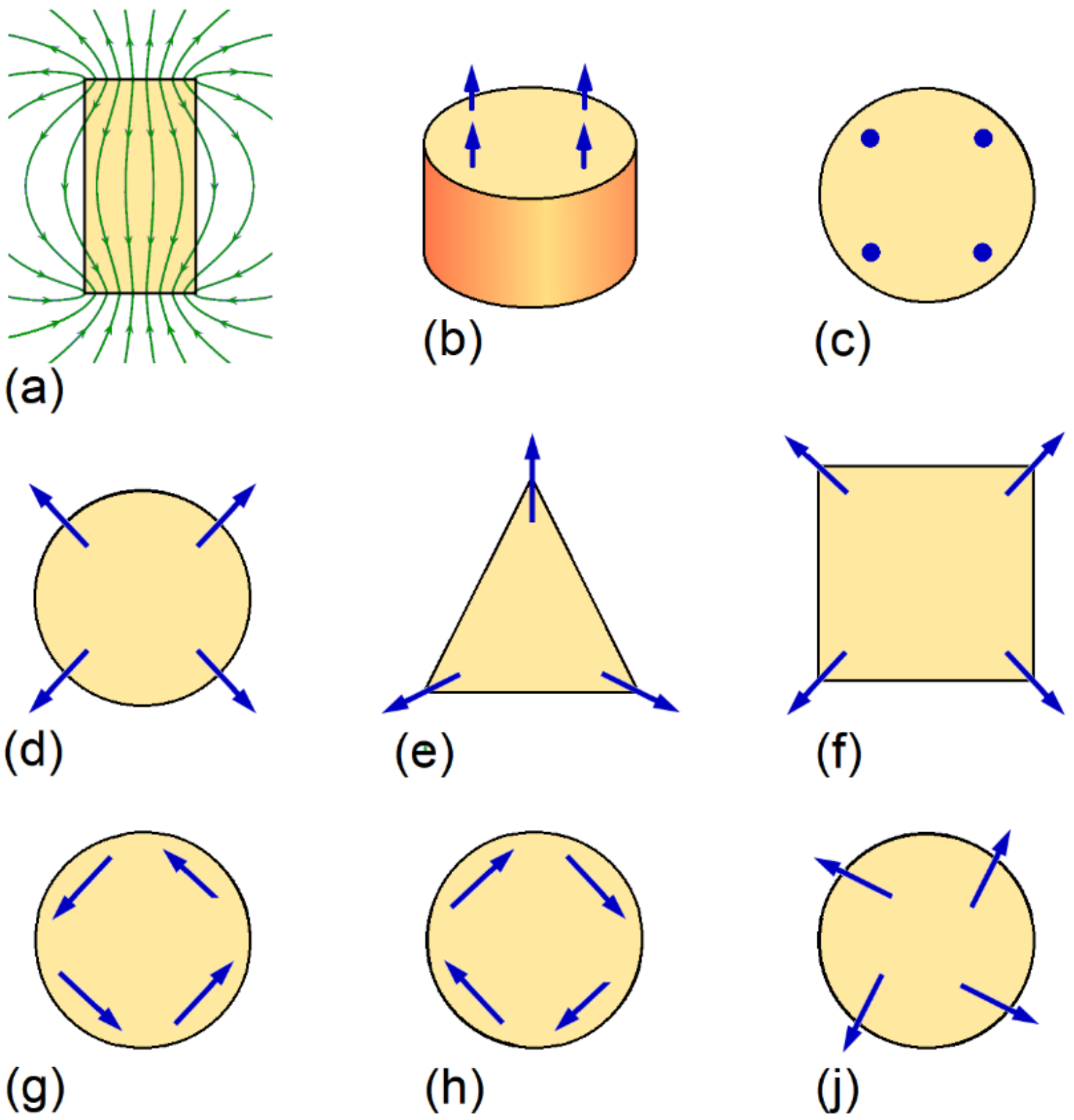

**Fig. 2**. Competing spin structures in small nanoparticles: (a) field distribution in the middle of a cylindrical nanoparticle, (b-c) uniform magnetization, (d) flower state in a cylindrical particle, (e-f) flower states in prismatic nanoparticles, (g-h) vortex states of opposite chirality and (j) mixed state. The spins $\boldsymbol{S}(\boldsymbol{r})$ are shown as blue arrows.

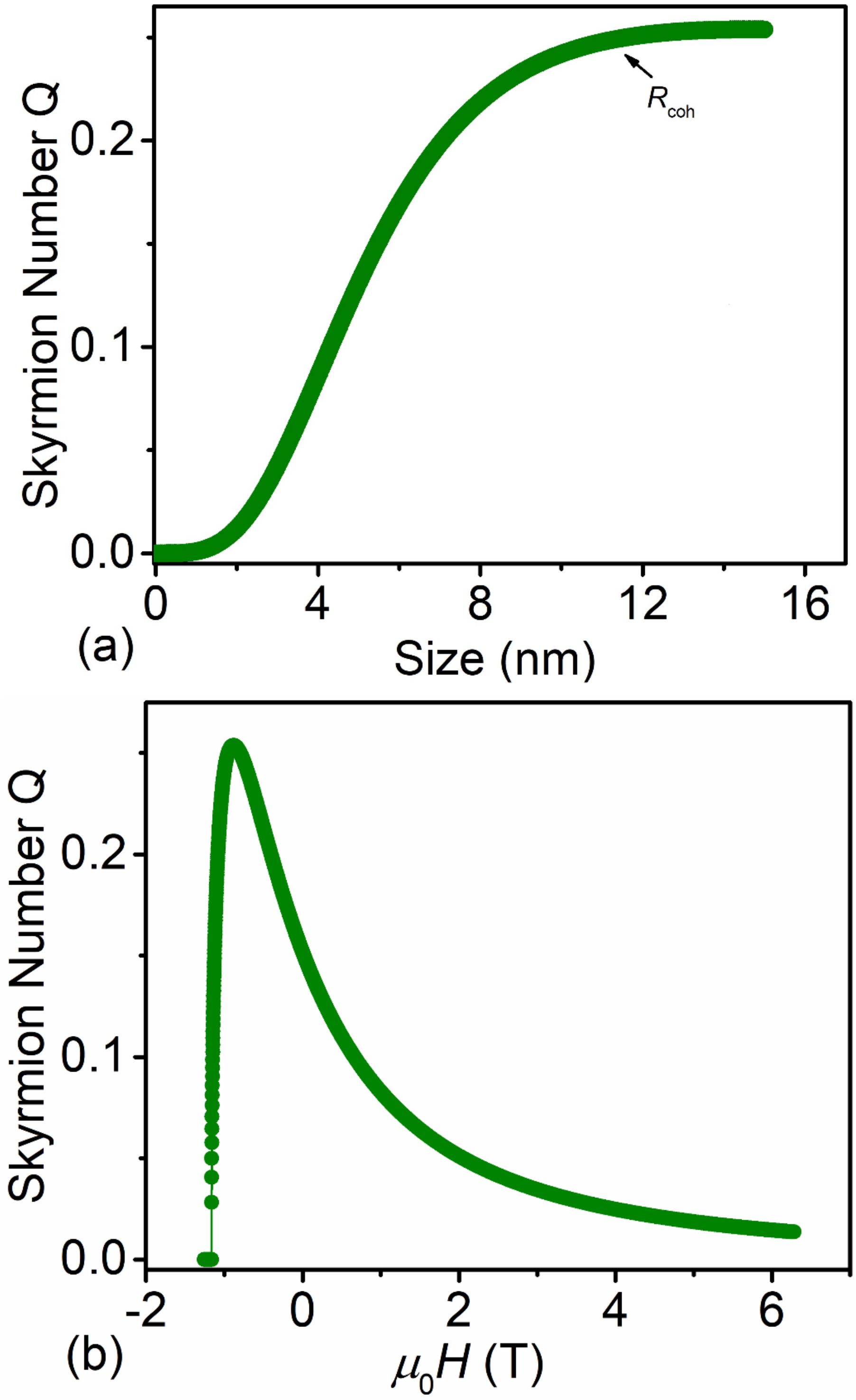

**Fig. 3.** Skyrmion number in very small particles: (a) particle-size dependence in the absence of an external magnetic field and (b) field dependence. In (a) at the maximum value where the skyrmion number is constant represent i.e., $R = R_{coh}$ where flower state vanishes, and curling mode appears.

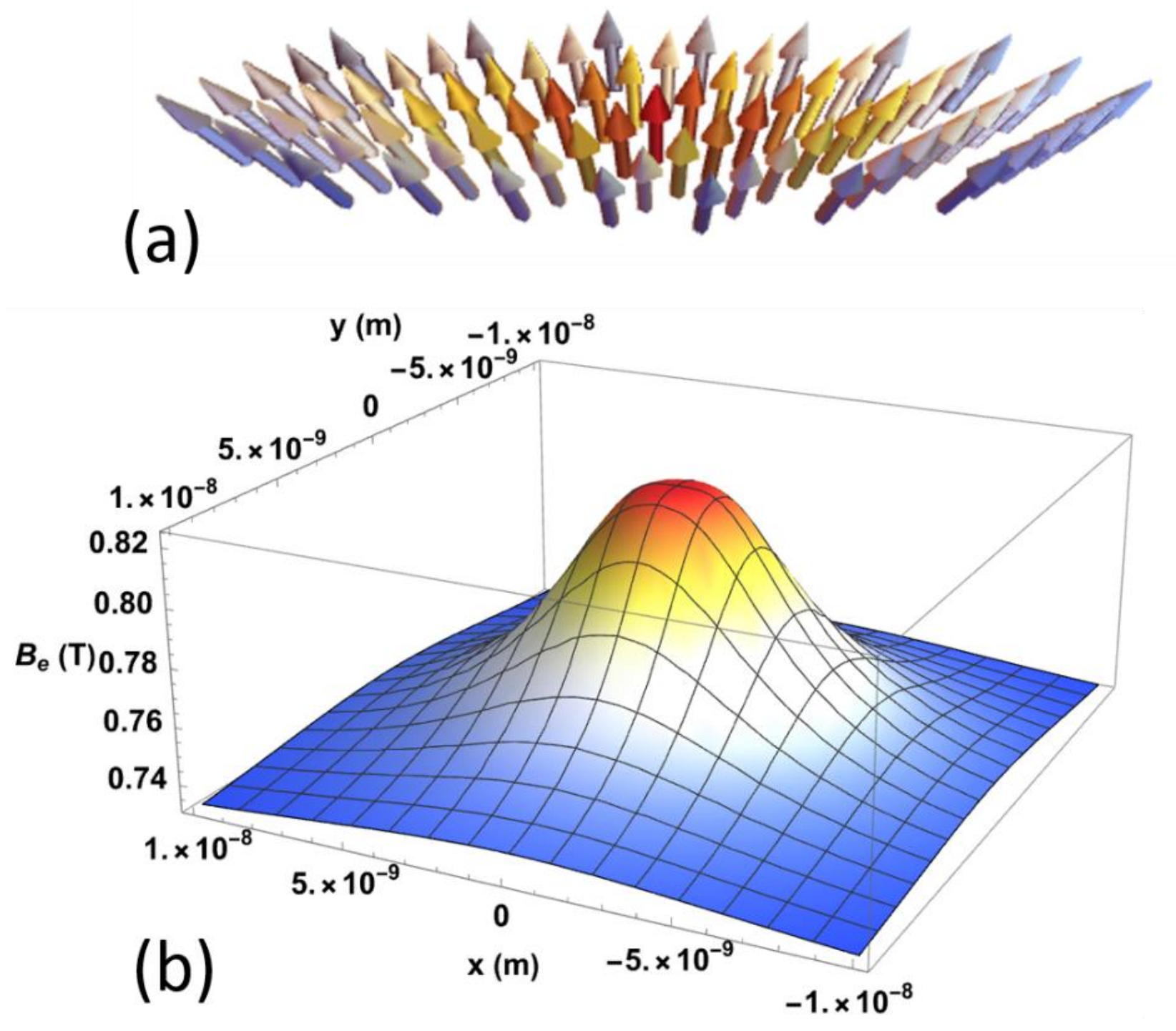

**Fig. 4.** (a) Vectro3d plot for flower state, (b) emergent magnetic field due to flower state in the cylinder of radius 10 nm.

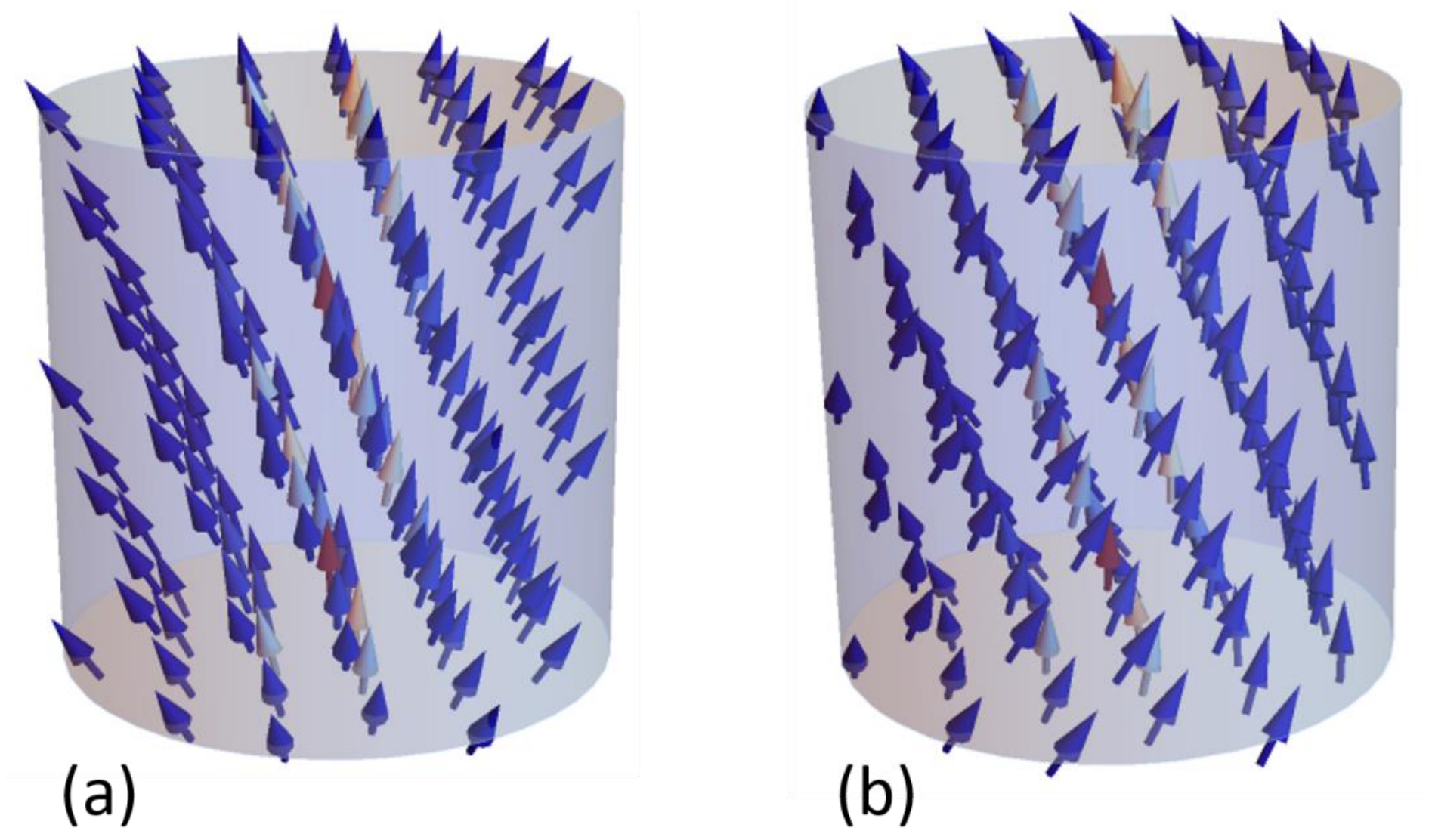

**Fig. 5.** Vector 3d plots in the cylinder using Eq. 3: (a) flower state with $\Phi(\boldsymbol{r})= \varphi$ and (b) curling state with $\Phi(\boldsymbol{r})= \varphi\pm\pi/2$.

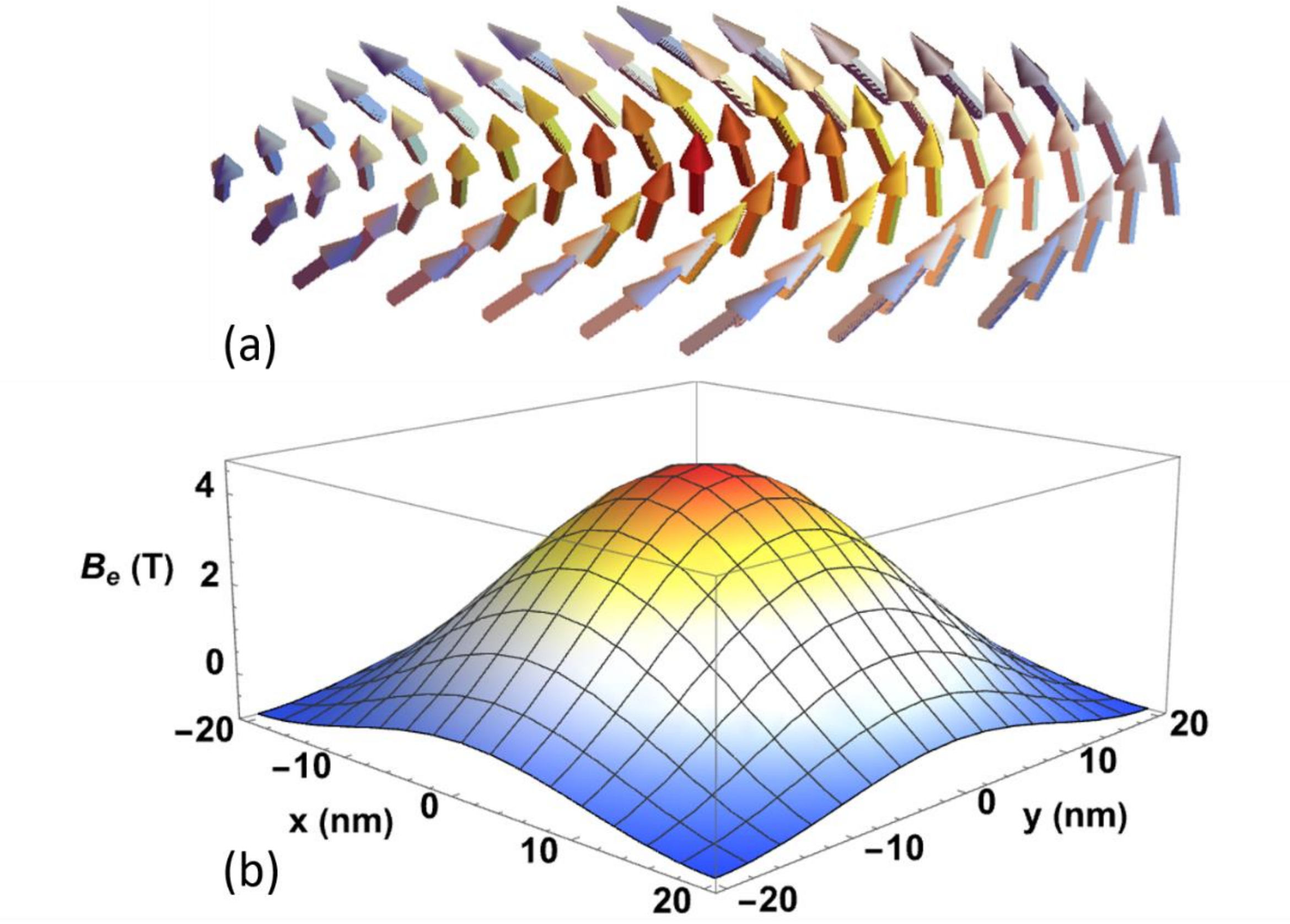

**Fig. 6.** (a) Vectro3d plot for curling mode using exact curling mode, (b) emergent magnetic field due to curling mode in the cylinder of radius 20 nm.

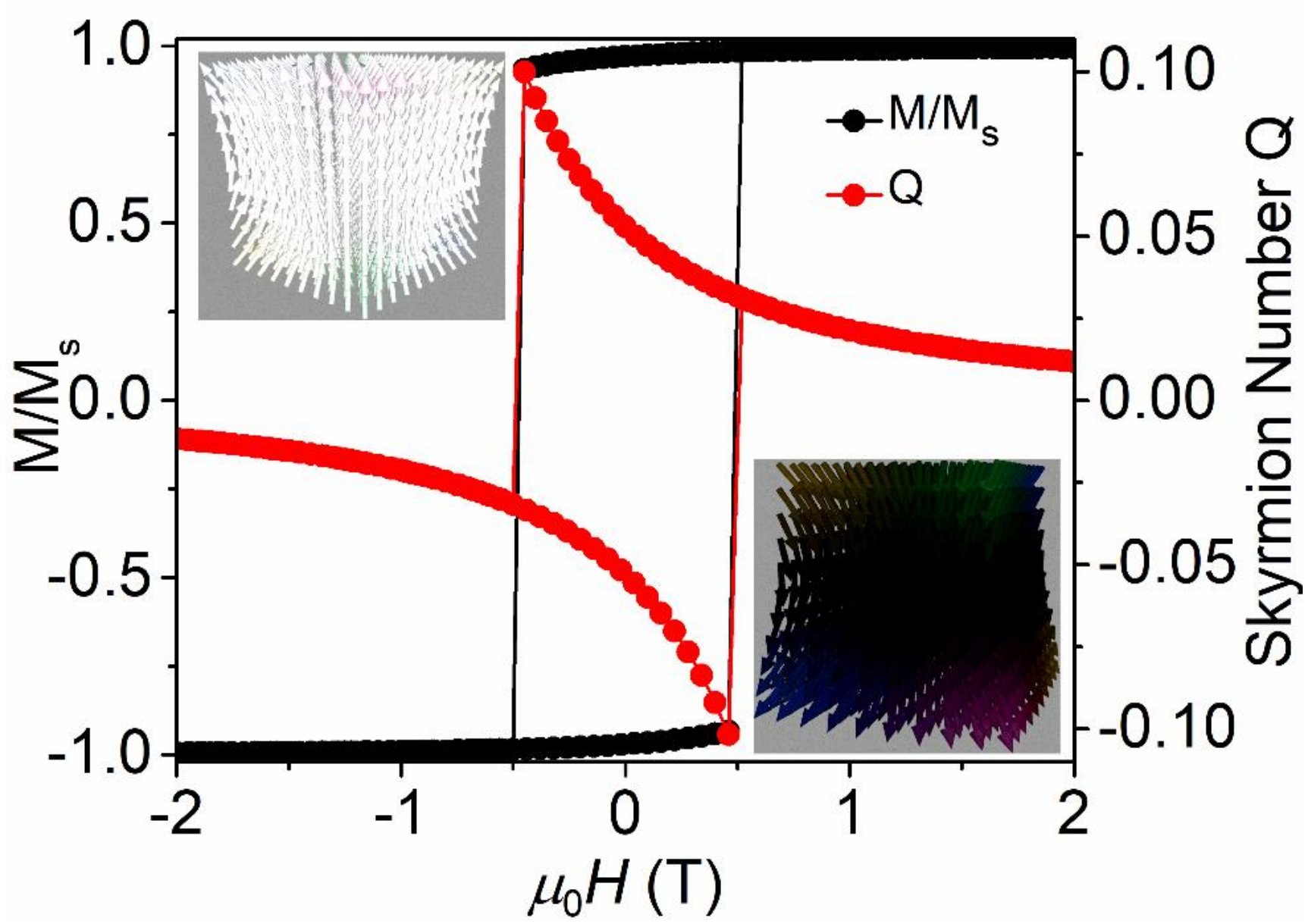

**Fig. 7.** Magnetization and Skyrmion number in a ferromagnetic cube with $L < L_{coh}$: Skyrmion number Q as a function magnetic field during magnetization reversal. The deviation from the saturation appears just before the reversal.

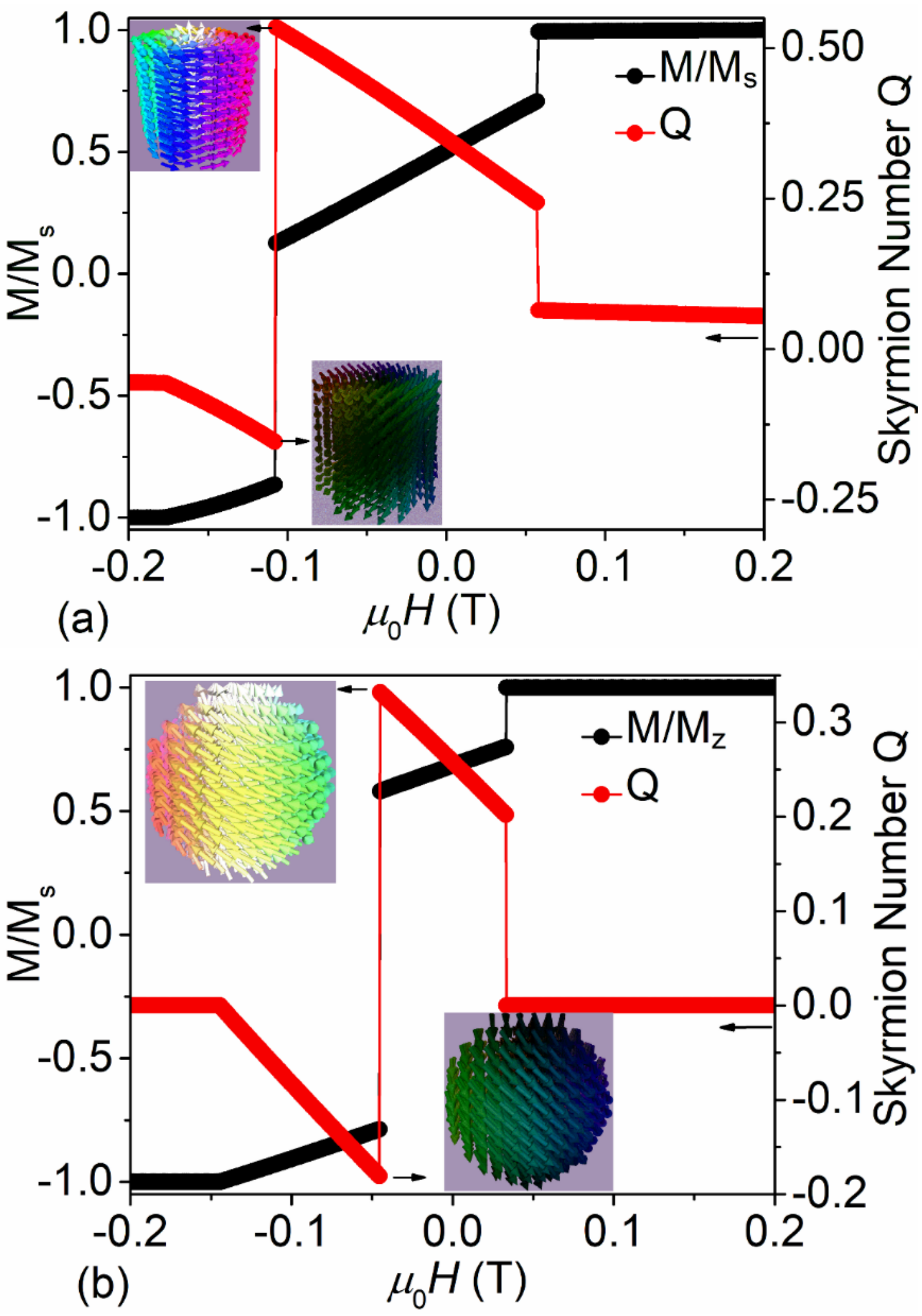

**Fig. 8.** Magnetization and Skyrmion number in ferromagnetic (a) cylinder (b) sphere for $R > R_{coh}$.

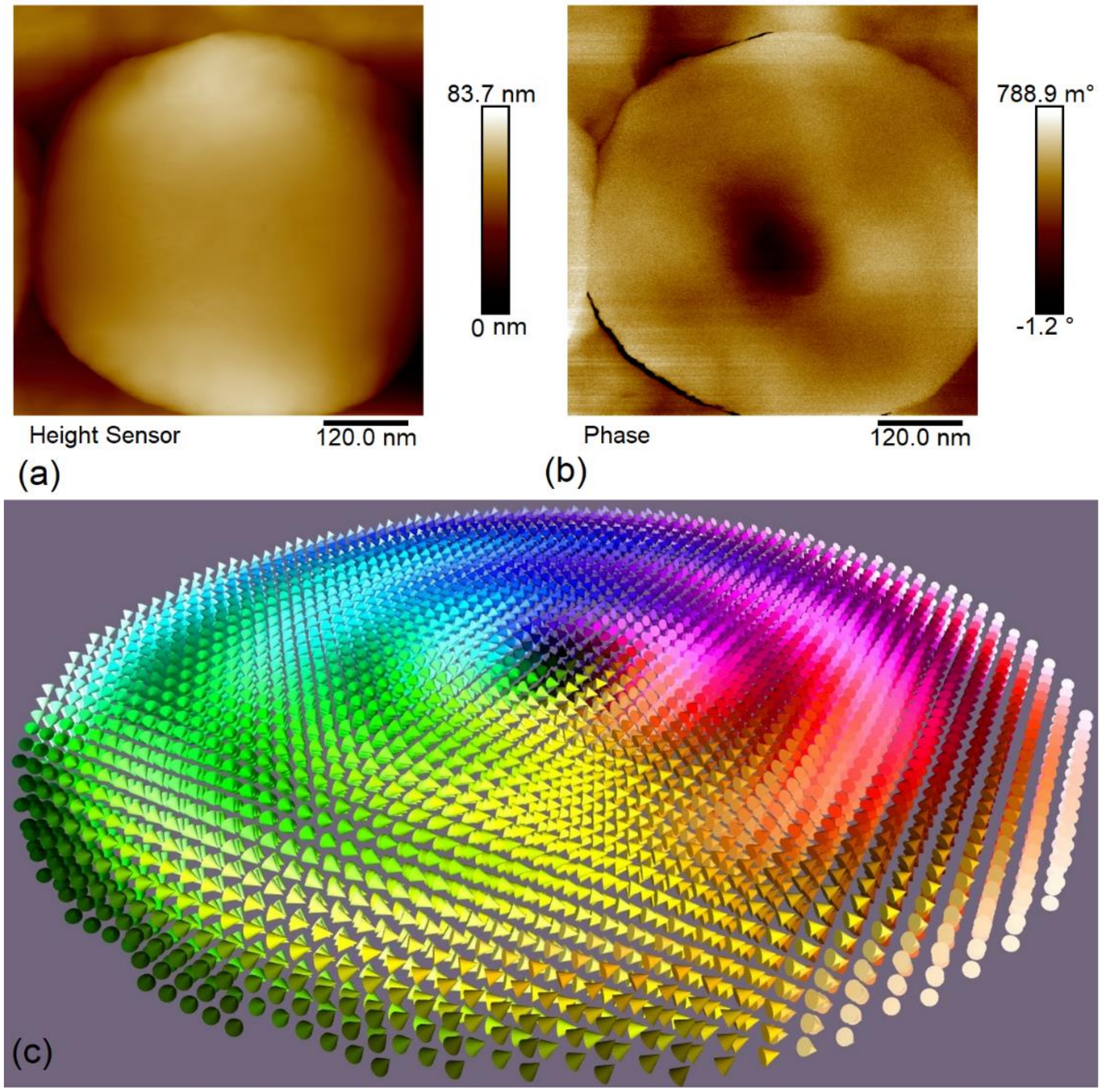

**Fig. 9.** (a) AFM for a pattern of circular nanodots, (b) MFM of circular nanodots showing magnetic vortex core observation in circular Dots, and (c) Micromagnetic simulation of Co nanodot of 500 nm diameter showing the presence of vortex state.

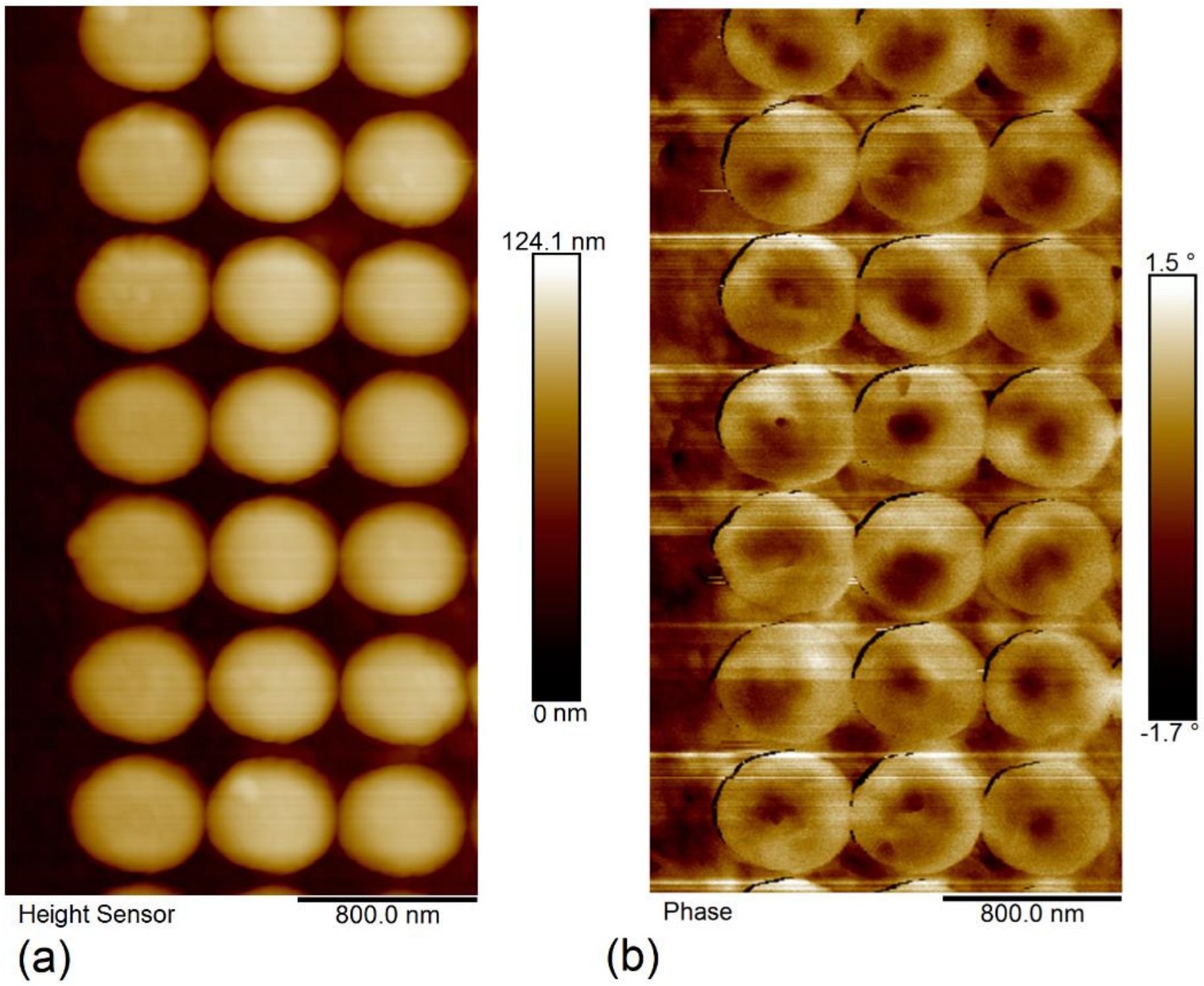

**Fig. 10.** (a) AFM for a pattern of circular nanodots (b) MFM of circular nanodots showing the presence of single domain vortex states.

# Berry Curvature and Topological Hall Effect in Magnetic Nanoparticles

Ahsan Ullah*, Balamurugan Balasubramanian, Bibek Tiwari, Bharat Giri , David J. Sellmyer‡, Ralph Skomski‡, Xiaoshan Xu
*Department of Physics & Astronomy and Nebraska Center for Materials and Nanoscience, University of Nebraska, Lincoln, NE 68588*
*E-mail: aullah@huskers.unl.edu
‡Deceased

## Calculations

$$\mathscr{E} = \int \left\{ A\left[\nabla\left(\frac{\boldsymbol{M}}{M_s}\right)\right]^2 - K_1 \frac{(\boldsymbol{n}.\boldsymbol{M})^2}{M_s^{\,2}} - \mu_o\, \boldsymbol{M}.\boldsymbol{H} - \frac{\mu_o}{2} \boldsymbol{M}.\boldsymbol{H}_d(\boldsymbol{M}) \right\} dV \qquad \text{(S1)}$$

$$\boldsymbol{M}(\boldsymbol{r}) \;= M_s\boldsymbol{S} = M_s \left(S_z\boldsymbol{e}_z \;+ \boldsymbol{S}_{\mathbf{xy}}\,(\boldsymbol{r})\right) \qquad \text{(S2)}$$

$$\boldsymbol{S}(\boldsymbol{r}) = \; \sin\Theta(\boldsymbol{r}) \cos\Phi(\boldsymbol{r})\, \boldsymbol{e}_x + \sin\Theta(\boldsymbol{r}) \cos\Phi(\boldsymbol{r})\, \boldsymbol{e}_y + \cos\Theta(\boldsymbol{r})\, \boldsymbol{e}_z,$$

$$\boldsymbol{S}_{\mathbf{xy}}(\boldsymbol{r}) = \sin\Theta(\boldsymbol{r}) \cos\Phi(\boldsymbol{r})\, \boldsymbol{e}_x + \sin\Theta(\boldsymbol{r}) \cos\Phi(\boldsymbol{r})\, \boldsymbol{e}_y \qquad \text{(S2)}$$

$$|\boldsymbol{S}_{\mathbf{xy}}(\boldsymbol{r})| = (\sin\Theta(\boldsymbol{r}))^2$$

$$\boldsymbol{M}(\boldsymbol{r}) \;= M_s \left(\sqrt{1 - S_{xy}{}^2(\boldsymbol{r})}\;\; \boldsymbol{e}_z \;+ \boldsymbol{S}_{\mathbf{xy}}(\boldsymbol{r})\right) \qquad \text{(S3)}$$

$$\boldsymbol{M}(\boldsymbol{r}) \;= M_s \left((1 - \frac{S_{xy}{}^2(\boldsymbol{r})}{2})\boldsymbol{e}_z \;+ \boldsymbol{S}_{\mathbf{xy}}(\boldsymbol{r})\right) \qquad \text{(S4)}$$

$$\nabla\, \boldsymbol{M}(\boldsymbol{r}) = M_s\nabla\boldsymbol{S}_{\mathbf{xy}}(\boldsymbol{r}) \qquad \text{(S5)}$$

$$\boldsymbol{n}.\boldsymbol{M} = M_s^2\,(1 - \frac{S_{xy}(\boldsymbol{r})}{2}^2) \qquad \text{(S6)}$$

$$\boldsymbol{M}.\boldsymbol{H} = M_s^2 H\,(1 - \frac{S_{xy}(\boldsymbol{r})}{2}^2) \qquad \text{(S7)}$$

$$\boldsymbol{M}.\boldsymbol{H}_d(\boldsymbol{M}) \;= -M_s^2\, \mathrm{D}\,(1 - S_{xy}(\boldsymbol{r})^2) \qquad \text{(S8)}$$

Aside from unimportant zero-point energy we have

$$\mathscr{E} \;=\; \int\left[A\,(\nabla\boldsymbol{S}_{xy}(\boldsymbol{r}))^2 + K\,(\boldsymbol{S}_{\mathbf{xy}}(\boldsymbol{r}))^2 + \frac{1}{2}\,\mu_o\,(H - \mathrm{D}\,M_s)\,M_s\,\boldsymbol{S}_{\mathbf{xy}}(\boldsymbol{r})^2]\right] dV \qquad \text{(S9)}$$

For small particles, $D = 0$, i.e. no magneto static self-interaction creation flux closure.

$$\mathscr{E} = \int \left[ A\,(\nabla \boldsymbol{S}_{xy}(\boldsymbol{r}))^2 + K\,(\boldsymbol{S}_{xy}(\boldsymbol{r}))^2 + \frac{1}{2}\,\mu_o\, H\, M_s\, \boldsymbol{S}_{xy}(\boldsymbol{r})\,^2] \right] dV \tag{S10}$$

In small nanoparticles, the stoner wolfarth assumes uniform magnetization but in reality, exchange interaction competes with other energy contributions such as magnetostatic interactions, etc. In very small nanoparticles the magnetostatic interaction is absent, but there is still a contribution of exchange interaction i.e. $A\nabla^2 = A/R^2$ where R is the radiuos of particle. So the volume averaging of eq. (S10) yields the approximation:

$$\mathscr{E}/\mathrm{V} = <\eta> = \left[ \frac{\mathrm{A}}{\mathrm{R}^2}\, \boldsymbol{S}_{\mathbf{xy}}(\boldsymbol{r})\,^2 + K\,(\boldsymbol{S}_{\mathbf{xy}}(\boldsymbol{r}))^2 + \frac{1}{2}\,\mu_o\, H\, M_s\, \boldsymbol{S}_{\mathbf{xy}}(\boldsymbol{r})\,^2 \right] \tag{S11}$$

For a moderately large particle with edges, the nanoparticle exhibits a flower state to minimize the energy. The flower state is given by $\boldsymbol{S} = \sin\Theta \boldsymbol{e}_R + S_z\, \boldsymbol{e}_z = S_{xy}(\boldsymbol{r})\, \boldsymbol{e}_R + S_z\, \boldsymbol{e}_z$ , and carries only a little anisotropy energy. Note that we used $\Phi(\boldsymbol{r}) = \varphi$ for flower state and radial component $\boldsymbol{e}_R = \cos\varphi\, \boldsymbol{e}_x + \sin\varphi\, \boldsymbol{e}_y$ and $S_{xy}(\boldsymbol{r}) = \sin\Theta(\boldsymbol{r})$.

The flower state causes a xy-component of the demagnetizing magnetostatic selfinteraction, $\mu_o \boldsymbol{M}(\mathrm{r})\cdot\boldsymbol{H}_d(\boldsymbol{M}(\mathrm{r}))/2$. Here $\boldsymbol{H}_d$ contains a uniform demagnetizing-field contribution $H_u = -H_u \boldsymbol{e}_z$ and a non-uniform or flower-state contribution perpendicular to $\boldsymbol{e}_z$. For particle $R < R_{coh}$, the flux closure due to magnetostatic-selfinteraction is absent. The non-uniform or flower-state contribution self-interaction is perpendicular to $\mathbf{e}_z$ i.e. $|\, S_{xy}(\boldsymbol{r})| = \sin\Theta$, given by additional demagnetization term as:

$$\boldsymbol{H}_F.\boldsymbol{S} = M_s\, H_F \sin\Theta \tag{S12}$$

Hence Eq. (S11) with flower state energy is given by:

$$<\eta> = \left[ \frac{\mathrm{A}}{\mathrm{R}^2} \sin\Theta\,^2 + K\sin\Theta\,^2 + \frac{1}{2}\,\mu_o\, H\, M_s \sin\Theta\,^2 - \mu_o\, M_s\, H_F \sin\Theta \right] \tag{S13}$$

As the size of the particle increases, flux closure due to magneto-static self-interaction gives rise to curling mode,

$$\boldsymbol{M}(\boldsymbol{r}) = M_s \left(S_z\mathbf{e}_z + S_{xy}(r)\mathbf{e}_P\right) = \left(S_z\mathbf{e}_z - S_{xy}(r)\sin\phi\, \boldsymbol{e}_x + S_{xy}(r)\cos\phi\, \boldsymbol{e}_y\right) \quad \text{(S14)}$$

And the energy equation is given by eq. (S9).

$$\mathscr{E} = \int \left[ A\,(\nabla \boldsymbol{S}_{xy}(\boldsymbol{r}))^2 + K\,(\boldsymbol{S}_{xy}(\boldsymbol{r}))^2 + \frac{1}{2}\mu_o\,(H - \mathrm{D}\,M_s)\,M_s\,\boldsymbol{S}_{xy}(\boldsymbol{r})^2] \right] \mathrm{d}V \quad \text{(S14)}$$

The energy

$$\mathscr{E} = \int \eta\, \mathrm{d}V \quad \text{(S15)}$$

is minimized by Euler Lagrange equation such that

$$\left[ \frac{\delta \eta}{\delta\, \boldsymbol{S}_{xy}\,(\mathrm{r})} = -\nabla\left(\frac{\partial \eta}{\partial \nabla\, \boldsymbol{S}_{xy}\,(\mathrm{r})}\right) + \frac{\partial \eta}{\partial\, \boldsymbol{S}_{xy}\,(\mathrm{r})} \right] \quad \text{(S16)}$$

We get,

$$\left(2A\nabla^2 - 2K_1 - \mu_0 M_s H + \mu_0\, DM_s^{\,2}\right) \boldsymbol{S}_{xy}(r) = 0 \quad \text{(S17)}$$

Where we used $\nabla(A\nabla\, \boldsymbol{S}_{xy}) \approx A\, \nabla^2\, \boldsymbol{S}_{xy}$. This eq. S17 is the same as used by Aharoni in Ref [10], i.e. Eq. 9.1.4 and 9.1.5, derived with the use of the calculus of variations.

The mode, which appears during magnetization reversal, are either coherent, flower or curling mode. The coherent mode is given by $\boldsymbol{S}(\mathbf{r}) = \mathrm{S}(\mathrm{r})\mathbf{e}_m$ with $\mathbf{e}_z.\mathbf{e}_m = 1$, gives zero skyrmion density. The flower state $\boldsymbol{S} = \sin\theta\, \boldsymbol{e}_R + S_z\, \boldsymbol{e}_z$ , has finte skyrmion density which increased by increasing the size. The curling mode $\boldsymbol{S} = \left(\Theta(\rho)\cos(\phi + \phi_o)\, \boldsymbol{e}_x + \Theta(\rho)\sin(\phi + \phi_o)\, \boldsymbol{e}_y + (1 - \Theta(\rho)^2)\mathbf{e}_z\right)$ has skyrmion number greater then flower state. During the magnetization reversal at nucleation field the local instabilities in the magnetization start. In small particles at the nucleation field the curling mode appears. For cylinder and sphere if radius $R > R_{coh}$ then at the nucleation field curling mode appears. When the field is applied parallel to the cylinder, the following physical scenario is realized. Starting from H = + infinity and reducing the field, one initially has $M_z = M_s$ (full saturation). This state persists through H = 0 until $H = |H_n|$, where $H_n$ is the nucleation field of the curling mode. At this point, the curling mode becomes more stable than the remanent mode $S_{x/y}$ = 0. For wide range of materials usually $R_{coh}$ = 10 nm. The curling mode is defined for small perpendicular magnetization components $\boldsymbol{S_{xy}}$ only, meaning that $M_z \sim M_s$.

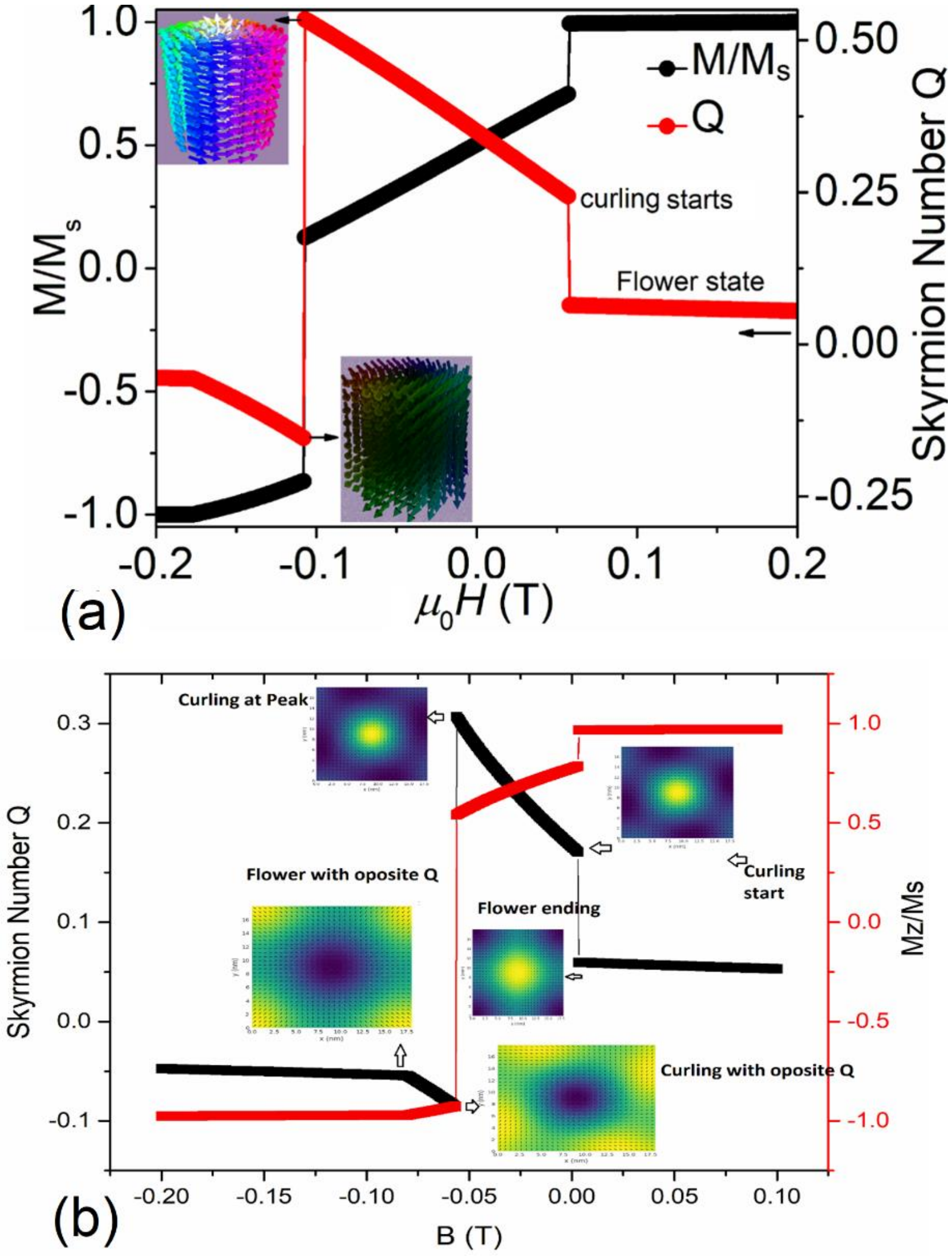

**Fig. S1. Bird view**: Reversal in magnetic nanodot with radius/length greater than coherence radius/length. (a) Cylinder (b) Cube. At a high magnetic field, the spins align in the direction of the external field giving a finite skyrmion number due to flower state. But at the nucleation field due to flux closure, the skyrmion number jumps to a higher value representing the flux closure state.

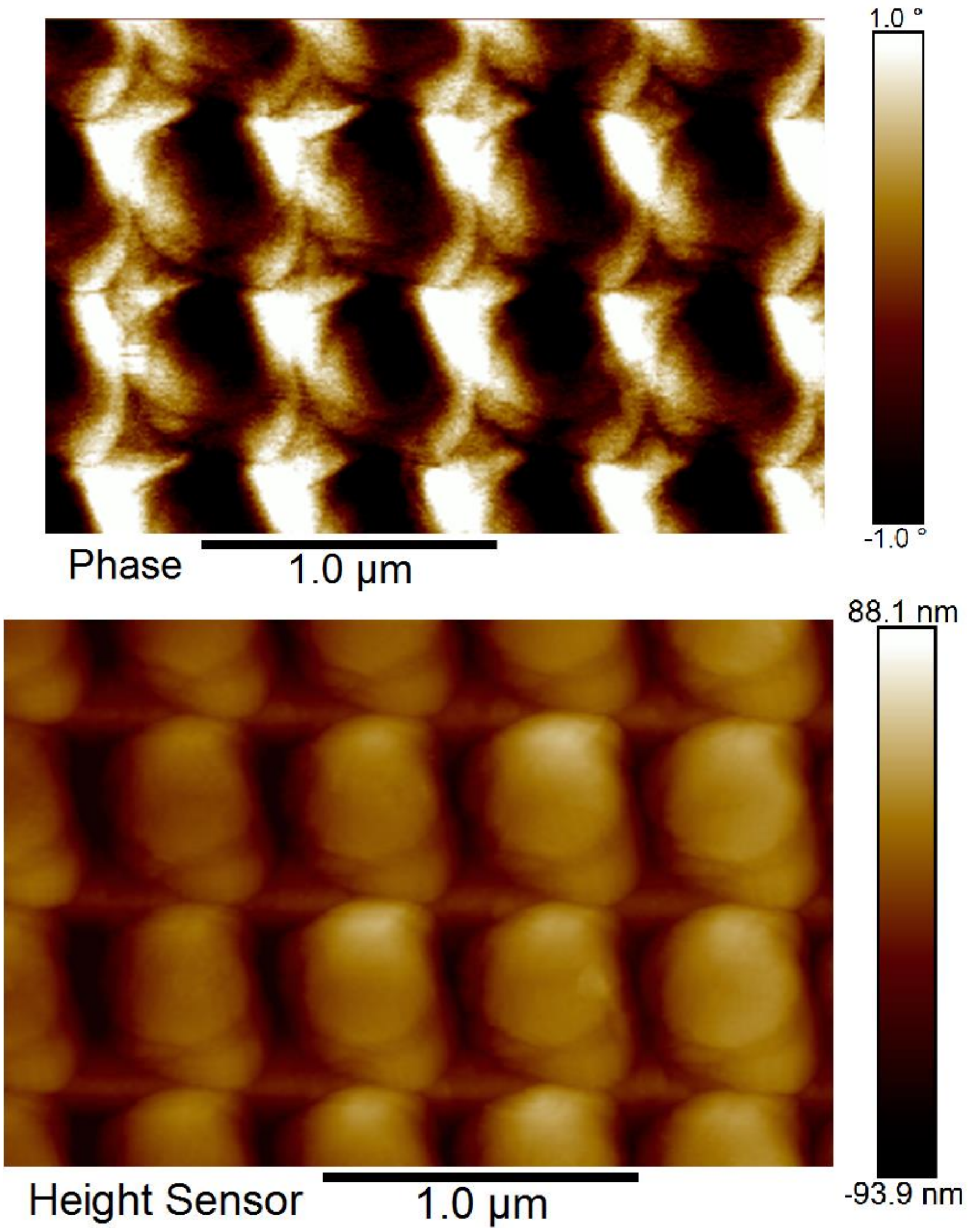

**Fig. S2.** Magnetic nanodots arrays in the presence of magnetic field showing magnetic vortex state desapear and showing the presence of tilted spin at the edges.